# Objects and Goals Extraction from Semantic Networks : Applications of Fuzzy SetS Theory

Mohamed Nazih OMRI

*Department of Mathematical and Data Processing,*
*Preparatory Institute of Engineering Studies of Monastir*
*Kairouan Road, 5019 Monastir, Tunisia*
Email: Nazih.omri@ipeim.rnu.tn

**Abstract.** In this paper we present a short survey of fuzzy and Semantic approaches to Knowledge Extraction. The goal of such approaches is to define flexible Knowledge Extraction Systems able to deal with the inherent vagueness and uncertainty of the Extraction process. In this survey we address if and how some approaches met their goal.

## 1. Introduction

The goal of a Knowledge Extraction System (KES) is to extract knowledge considered pertinent to a user's request, expressed in the natural language. The effectiveness of a KES is measured through parameters, which reflect the ability of the system to accomplish such goal. However, the nature of the goal is not deterministic, since uncertainty and vagueness are present in many different parts of the extraction process. The user's expression of his/her knowledge needs in a request is uncertain and often vague, the representation of an Object and or a Goal informative content is uncertain, and so is the process by which a request representation is matched to an Object representation. The effectiveness of a KES is therefore crucially related to the system's capability to deal with the vagueness and uncertainty of the Extraction process. Commercially available KESs generally ignore these aspects; they oversimplify both the representation of the Objects' content and the user-system interaction. A great deal of research in KE has aimed at modeling the vagueness and uncertainty, which invariably characterize the management of knowledge. A first glans of approaches is based on methods of analysis of natural language [1]. The main limitation of these methods is the level of deepness of the analysis of the language, and their consequent range of applicability: a satisfying interpretation of the Objects' meaning needs a too large number of decision rules even in narrow application domains. A second glans of approaches is more general: their objective is to define Extraction models, which deal with imprecision, and uncertainty independently on the application domain. The most long-standing set of approaches belonging to this class goes under the name of Probabilistic Information Retrieval (IR) [2]. There is another set of approaches receiving increasing interest that aims at applying techniques for dealing with vagueness and uncertainty. This set of approaches goes under the name of *Knowledge Extraction.*

In this chapter we will review some of the approaches to Knowledge Extraction, in particular the approaches that make use of Fuzzy Set Theory and Semantic Networks. The remainder of this chapter is structured as follows: in section 2. we describe the structure of fuzzy Goals and fuzzy Objects used in the semantic network. In section 3. we present an introduction to the KE problem. In section 3.1 we give an overview of the main classical KE models, while in section 3.2 we explain the Object of Knowledge Extraction. The core of the chapter is in properties 4. where a number of models of Knowledge Extraction based on Fuzzy Set Theory. In section 5. we draw the conclusions of our survey and outline future works of research.

## 2. Fuzzy Goals and Fuzzy Objects

In fuzzy logic, there has been a lot of research into the notion of knowledge extraction. However, so far, researchers have mainly analysed the information retreival from document, while we need a somewhat more sophisticated notion: goal extraction from fuzzy sets to fuzzy goals and objects in a semantic Network. To extend definitions of knowledge (goal) extraction from fuzzy sets to fuzzy goals and objects, we must first analyse how fuzzy objects and goals can be described in terms of fuzzy sets [3, 4].

How does a user formulate his or her gaols? For example, how do we describe a goal when we look for a house to buy? A natural goal is to have a house not too far away from work, not too expensive, in a nice neighborhood, etc. In general, to describe a goals:

- we list *attributes* (in the above example, distance, cost, and neighborhood quality), and
- we list the desired (fuzzy) value $A_1,…,A_n$ of these attributes (in the above example, these values are, correspondingly, "not too far", "not too expensive", and "nice").

Each of the fuzzy values like "not too far" can be represented, in a natural way, as a fuzzy set.

Similarly, an object can be described if we list the attributes and corresponding values. At first glance, it may seem that from this viewpoint, a description of an object is very much alike the description of a goal, but there is a difference.

## 3. Knowledge Extraction

*Knowledge Extraction* is a branch of Computing Science that aims at storing and allowing fast access to a large amount of knowledge. A Knowledge *Extraction System i*s a computing tool, which represents and stores knowledge to be automatically extracted for future use. Most actual KE systems store and enable the Extraction of only textual knowledge or Objects. However, this is not an easy task, it must be noticed that often the sets of Objects a KES has to deal with contain several thousands or sometimes millions of Objects and/or Goals.

A user accesses the KES by submitting a request; the KES then tries to extract all Objects and/or Goals that are "relevant" to the request. To this purpose, in a preliminary phase, the Objects contained in the Semantic Network (SN) [5, 6] are analyzed to provide a formal representation of their contents: this process is known as "indexing". Once an Object has been analyzed a surrogate describing the Object is stored in an index, while the Object itself is also stored in the SN. To express some knowledge needs a user formulates a request, in the system's request language. The request is matched against entries in the index in order to determine which Objects are relevant to the user. In response to a request, a KES can provide either an exact answer or a ranking of Objects that appear likely to contain knowledge relevant to the request. The result depends on the formal model adopted by the system. As it will be explained in next section, the Boolean model produces an exact answer, while others, most advanced models apply a partial matching mechanism, which produces a ranking of the extracted Objects so that the most likely to be relevant are presented to the user first. In some KESs requests are expressed in natural language and to be processed by the system they are passed through a request processor, which breaks them into their constituents words.

### 3.1. *Knowledge extraction models*

In the KE literature different models have been proposed.

The *Boolean model* is still the one most commonly used in commercial KE systems. It is based on mathematical set theory. Here Objects are represented as sets of index goals, whose role cannot be differentiated to express the knowledge content. A request is a logical formula made up of index goals and logical connectives (e.g. AND, OR, NOT). An Object is considered relevant and extracted by the KES if it satisfies the logical formula representing the request.

The *Vector Space model* [7] is based on a spatial interpretation of both Objects and requests. Here an improvement of the Objects representation over the Boolean model is obtained by associating with each index goal a numeric value, called the index goal weight, which expresses the variable degree of significance that the goal has in synthesizing the knowledge content of the Object. Similarity measures between Object and request representation are then used to evaluate an Object's relevance with regards to a request.

The *Probabilistic model* [8] ranks Objects in decreasing order of their evaluated probability of relevance to a user's knowledge need. Past and present research has made much use of formal theories of probability and of statistics in order to evaluate, or at least estimate, the probability of relevance. Without going into the details of any of the large number of probabilistic models of KE that have been proposed in the literature (for a survey see [2]), if we assume that an Object is either relevant or not relevant to a request, the task of a probabilistic IR system is to rank Objects according to their estimated probability of being relevant, i.e. *P(R I q, d)*. Probabilistic relevance models base this estimation on evidence about which Objects are relevant to a given request. The problem of estimating the probability of relevance for every Object in the set is difficult because of the large number of variables involved in the representation of Objects in comparison to the small amount of Object relevance knowledge available.

The above-mentioned models are the most studied ones. However, a large number of other models have been investigated and used in prototypical KES.

### 3.2. *Knowledge extraction*

In recent years big efforts have been devoted to the attempt to improve the performance of KE systems and research has explored many different directions trying to use with profits results achieved in other areas. In this paper we will survey the application to KE of two theories that have been used in Artificial Intelligence for quite some time: Fuzzy set theory and SN theory. The use of fuzzy set or SN techniques in KE has been recently refered to as *Knowledge Extraction* in analogy with the areas called Computing and Information Retreival.

*Fuzzy set theory* [9] is a formal framework well suited to model vagueness: in information retreival it has been successfully employed at several levels [10, 11, 33], in particular for the definition of a superstructure of the Boolean model, with the appealing consequence that existing Boolean KESs can be improved without redesigning them completely [14, 15, 17]. Through these extensions the gradual nature of relevance of Objects to user requests can be modelled.

A different approach is based on the application of the SN *theory* to KE. Semantic Networks have been used in this context to design and implement KESs that are able to adapt to the characteristics of the KE environment, and in particular to the user's interpretation of relevance.

In this chapter we will review the applications of fuzzy set theory and Semantic networks to KE.

## 4. Application of Fuzzy Set Theory to Knowledge Extraction

In order to increase the flexibility of KESs some approaches based on the application of fuzzy set theory have been defined. A fuzzy set allows the characterisation of its elements by means of the concept of "graduality"; this concept supports a more accurate description of a class of elements when a sharp boundary of membership cannot be naturally devised [9]. The entities involved in a KES are well suited to be formalised within this formal framework with the aim of capturing their inherent vagueness.

In the following properties a short survey of these approaches is presented. Other applications of fuzzy set theory have concerned the definition of knowledge-based models of KE, and the definition of fuzzy measures for evaluating the effectiveness of KESs, in goals of recall and precision. These approaches are not described here, but the interested reader can read, among others, the contributions in [16, 9, 10] relating to the former applications, and the contribution in [18] relating to the latter applications.

### 4.1. *Fuzzy object representations*

A first natural extension of the Boolean model is to represent an Object as a fuzzy set of goals, thus making the description of the Object's knowledge contents more accurate. For each goal associated with an Object a numeric weight is specified (the membership degree), which expresses the level of concern of the goal with respect to the knowledge contained in the Object; formaily, the function defining the relation between Objects and Goals is defined as: $F: O \times G \rightarrow [0,1]$.

An Object is then represented as a fuzzy set of goals, $\{\mu(g)/g\}$, in which $\mu(g) = F(o, g)$. The fuzzy Object representation is then based on the definition of a weighted indexing function, which for each pair goal-Object produces a numeric value, the so-called index goal weight. Like in the vector space model and in the probabilistic model, the use of index goal weights makes the Extraction mechanism able to rank Objects in decreasing order of their relevance to the user request, the relevance being expressed by a numeric score, the so called Extraction Status Value. The quality of the Extraction results strongly depends on the definition of the adopted weighting function; the original proposal in the literature defined this function by means of a count of goal occurrences in the Object and in the whole SN [7].

This definition, however, does not take into account that the knowledge in Objects is often structured: an Object in the SN, for example, is organised in properties as *is-a, kind-of, Goals, similar Objects,* etc. In such a configuration of an Object, a single occurrence of a goal in the Goals property suggests that the Object is fully concerned with the Object expressed by the goal, while a single occurrence in the similar Objects property indicates that the Object refers to other Objects in relation with that Object. It is evident that the knowledge carried by a goal occurrence depends on the property where it is located. The properties of an Object may assume a different importance on the basis of users' needs. For example, when looking for Objects witch are son of a certain class, the most important property to analyze is *the kind-of* property, while when looking for Objects on witch a given goal can be applied, the Goals property should be first analysed.

To face with this problem we have proposed a fuzzy representation of structured Objects, which can be biased by user's interpretation [5]. The significance of a goal g in a given Object o is computed by first evaluating the significance of g in each of the n properties; this is done by means of the application of a function $F_{pi}$ which has to be defined for property pi *($F_{pi}$ (o, g),* denotes the significance degree of goal g in property pi of Object o). Moreover, with each Object's property the user can associate a numeric importance in the set [0, 1], which is used in the aggregation phase to emphasise the role of the *$F_{pi}$ (g, o) s* of important properties with respect to those of less important ones. The significance degrees *$F_{p1}$ (o, g),...,, $F_{pn}$(o, g)* are then aggregated by means of a function, which can be selected by the user among a predefined set of linguistic quantifiers: *all*, *at least one, or almost k.* The linguistic quantifier indicates the number of Objects' properties in which a goal must be present to be considered fully significant. This fuzzy representation of structured Objects has been implemented and evaluated, showing that it improves the effectiveness of a system with respect to the use of the traditional fuzzy representation [13].

Based on this assumption, an indexing function has been proposed, which computes the significance of a goal in an Object by taking into account the different role of goal occurrences according to the importance of tags in which they appear. The significance degree of a goal in an Object is obtained by first computing the significance degrees of the goal inside the different tags, and then by aggregating these values taking into account the different importance of tags. A ranking of tags based on their importance allows assigning a different numeric weight to each tag, and consequently the contribution of a word occurring in a tag is modulated by this importance weight.

### 4.2. *Fuzzy extensions of the request language*

A request formulated through the Boolean language can be seen as the specification of a set of selection criteria connected through the Boolean operators AND and OR. A selection criterion is the elementary block for requesting information; in the Boolean language it is constituted by a term, which is selected by a user as the carrier of the concept synthesising the information he/she is looking for. To express more structured requests the elementary selection criteria can be put in relation by means of the aggregation operators. The fuzzy extensions of the Boolean request language proposed in the literature have concerned different levels. A first level of extension was directed to extend the selection criteria. This is done by defining selection criteria as pairs *term-importance weight,* in which the weight specifies the importance of the search term in the desired documents. In this paper, we define a selection creteria as pairs *goal-importance weight,* in which the weight specifies the importance of the search goal in the desired objects.

#### 4.2.1. *Numeric request weights*

Importance weights have been first formalised as numeric values, which specify a constraint to be satisfied by the fuzzy representation of objects in the SN. The function *f* matching a selection criterion *<g, w>* against an object *o* is defined as follows:
$$f: [0, 1] \times [0, 1] \rightarrow [0, 1].$$
The value *f(F(o, g), w)* is the degree of satisfaction of the selection criterion *<g, w>* by object *o*. The satisfaction of the separability property by a KES makes it possible to evaluate a request from the bottom up, evaluating a given object against each weighted goal in the request and then combining those evaluations according to the request structure [11].

The nature of the constraint imposed by the weighted selection criterion depends on the semantics associated with the weight; in the literature different semantics for request weights have been proposed. The weight can be interpreted as an importance weight, as a threshold, or as description of an "ideal" object. The semantics adopted for request weights implies a different definition of function *f*.

Some authors, among which *Radecki, Bookatein, Yager* have interpreted request weights as indicators of the relative importance among terms in a request [21, 24, 25]. The problem with this semantics however is related with its dependence on the type of the aggregation operator, which connects pairs of selection criteria. When using an AND, for example, a very small value of *w* for one of the two terms will dominate the *min* function and force a decision based on the least important (smallest *w*) term, which is just the opposite of what is desired by the user [10]. To overcome this problem the proposed definitions of the g function violate the property of separability [11].

Another semantics for request weights is to interpret them as thresholds that indicate the minimum acceptance level for the goal significance degree in an object, to be selected. *Radecki* first proposed the following simple definition for the *f* function, based of a λ-level meaning: $f(F(d, t), w) = 0$ for $\lambda \leq w$, while $f(F(d, t), w) = F(d,t)$ for $\lambda > w$ [21].

*Knaft and Buell* proposed, by allowing some small partial credit for a document whose $F(d, t)$ value is less than the threshold [10]. For $F(d, t) < w$, the function *f* returns a value varying *as* the percent satisfaction of *w* by the weight $F(d, t)$. For $F(d, t) < w$, the function g is a measure of the closeness of $F(d, t)$ to *w*, while for $F(d, t) > w$, g expresses the degree of over-satisfaction of the threshold *w*. A request <*t, w*> can be then interpreted as a request for the minimally acceptable document, which is the one having $F(d, t) = w$.

*Bordogna, Carrara and Pasi* [13] have interpreted request weights as specifications of ideal significance degrees. Based on work by *Gater and Kraft* [26], they have proposed a *f* function which measures the closeness of $F(d, t)$ to w: $f(F(d, t), w) = exp(K (F(d, t) - w)^2)$.

This inverse distance measure is symmetric in that documents with an $F(d, t)$ value greater than the *w* value are considered as those documents with an $F(d,t)$ value less than the w value.
Based on [13], *Kraft, Borlogna, and Pasi* [27] have proposed a closeness measure that allows for asymmetry.

### 4.2.2. *Request's object and goal weight*

The main limitation of numeric request weights is to force the user to quantify the qualitative and vague concept of importance. To make simpler and more natural the specification of some level of importance associated with the object and/or goal (o/g) in a request, linguistic weights have been formalised, such as *important, very important, fairly important*, etc.

To this aim *Borlogna and Pasi* [28] have defined a fuzzy retrieval model in which the linguistic descriptors are formalised within the framework of fuzzy set theory through linguistic variables [29]. A <*g, l*> pair identifies a qualitative selection criterion, where o/g is a object and/or goal and *l* is a value belonging to the object and/or goal set of the linguistic variable *Importance,* which has a base variable ranging over the set [0,1] (the admissible values of the indexing function *F).* Such a request language can be employed by any KES with a weighted object representation. To compute the degree of satisfaction of a pair <*g, l*> by a given object *o,* the compatibility of the index goal weight *F(o,g)* is evaluated with respect to the constraint imposed by the linguistic request weight *l*. An example of such a goal set is: *T (Importance) = {important, very important, notimportant, fairlyimportant,..}.*

The meaning of a linguistic value *l* is defined by means of a function $\mu_l$ which assesses the compatibility of the representation of objects, i.e. the *F (o, g)*s, with the linguistic goal *l*. The meanings of non-primary goals in *T(Importance)* are obtained by first defining the compatibility function associated with the primary goal important, $\mu_{importance}$, and then by modifying $\mu_{importance}$ according to the semantics of the hedges.

By considering the linguistic weight as a "fuzzification" of numeric weights, in [13] a simple procedure has been proposed to derive the semantics of the primary goal important from the semantics of numeric request weights. This procedure is based on the assumption that a linguistic request weight can be seen as the synthetic expression of a set of numeric weights; in other words when a user asks for objects in which the concept represented by goal *g* is *important,* he/she expresses a fuzzy concept on the term significance values (the *F(o, g)* values).

The function evaluating a pair <*g, l*> has then be defined as:

$$\mu_{importance}(F(o, g)) = MAX_{W \in [i,j]} g(F(o,g),w)$$

In which the definition of the g function depends on the semantics adopted for request weights. The two values, i and j, with i < j define the range of numeric values satisfying the linguistic constraint *important.* An example of function evaluating the linguistic weight important has been proposed in [27].

### 4.2.3. *Aggregation operators*

Aggregation operators are used to combine single selection criteria to express more complex request of information. The e function evaluating a Boolean query composed of n selection criteria is defined as: *h*: D x [0,1]$^n \rightarrow$ [0,1]. The arguments of function e are the degrees of satisfaction of the selection criteria (produced by the application of function *h*). When considering weighted selection criteria, the AND and OR connectives are interpreted as a *t-norm* and a *t-conorm* operators respectively. Usually the *min t-norm* and the *max t-conorm* are adopted. In the Boolean language the allowed connectives are the AND and the OR, which support crisp aggregations. For example, when evaluating a request composed by n selection criteria aggregated through the AND, the matching mechanism does not tolerate the unsatisfaction of a single criterion; this may cause the rejection of useful items. Although the fuzzy request expressions seen so far achieve a higher expressiveness than ordinary Boolean expressions, they do not reduce the complexity of Boolean logic. To face this problem, other extensions of the Boolean request language have concerned the definition of softer aggregation operators. To this aim new definitions of aggregation operators have been proposed; for example, *Salton, Fox and Wu* proposed a model based on a norm operator [30]. *Hayashi* [31], *Sanchez* [32] and *Paice* [33] consider "soft" Boolean operators weighted between the AND and the OR as a compromise. However, in these approaches different soft interpretations of the Boolean connectives in the same request are not supported.

### 4.3. *Fuzzy thesauri of goals*

Associative mechanisms are defined to enrich a KES so as to make it able to extract additional Objects, which are not indexed by the goals in a given request. Thesauri are an example of associative mechanisms, which exploit relations among goals, and are usually employed to select goals associated with request goals. Three main types of relation among goals are generally exploited; the relation *broader* goal (BG) is used to express that one goal has a more general meaning than the entry Goal. The relation *narrower* goal (NG) is the inverse relation; the relation related goal (RG) is defined to exploit synonyms or near-synonyms. Fuzzy thesauri have been defined in order to express the strength in the association between pairs of goals by analogy to pairs of terms [20, 34, 35]. The first works on fuzzy thesauri introduced the notion of fuzzy relations to represent associations between goals.

*Kohout, Kenavanou, and Bandler* [36] consider a synonym link to be a fuzzy binary relation defined in terms of fuzzy implication. The authors also define a narrower term link (where term $t_1$ is narrower than term $t_2$, so term $t_2$ is broader than term $t_1$) by means of fuzzy implication.

*We* introduced the following definition of a fuzzy relation. Let $O$ be a set of Objects. Each goal $g \in G$ corresponds to a fuzzy set of objects $h(g)$:

$$h(g) = \{\mu_g(o)/o \mid o \in O\} \quad (1)$$

in which $\mu_g(o)$ is the degree to which goal g is related to Object o. A measure $M$ is defined on all the possible fuzzy sets of Objects, which satisfies:

- $M(\phi) = 0$, $M(O) < \infty$,
- $M(A) < M(B)$ if $A \subseteq B$.

A typical example of M is the cardinality of a fuzzy set.

The similarity between two index goals, $(g_1, g_2) \in G$, is represented in a fuzzy thesaurus by the fuzzy relation *u* defined as:

$$u(g_1, g_2) = \frac{M(h(g_1) \cap h(g_2))}{M(h(g_1) \cup h(g_2))} \quad (2)$$

The fuzzy relation *v*, which represents grades of inclusion of a narrower goal $g_1$ in another (broader) goal $g_2$, is defined as:

$$v(g_1, g_2) = \frac{M(h(g_1) \cap h(g_2))}{M(h(g_1))} \quad (3)$$

By assuming *M* as the cardinality of a set, *u* and *v* are given as:

$$u(g_1, g_2) = \frac{\sum_k \min(\mu_{g_1}(c_k), \mu_{g_2}(c_k))}{\sum_k \max(\mu_{g_1}(c_k), \mu_{g_2}(c_k))} \quad (4)$$

$$v(g_1, g_2) = \frac{\sum_k \min(\mu_{g_1}(c_k), \mu_{g_2}(c_k))}{\sum_k \mu_{g_1}(c_k)} \qquad (5)$$

A fuzzy pseudo-thesaurus can be defined by replacing the set C in the definition of *h(g)* above with the set of Objects O, with the assumption that *h(g)* is the fuzzy set of Objects indexed by goal g. Thus, *h(g) = {(o, μ$_g$(o))/ o ∈ O}*, in which μ$_g$(o)= *F(o, g)*. F can be either a binary value or a value in [0,1], defining *a fuzzy* representation of Objects. The fuzzy RG and the fuzzy NG relations are defined *as:*

$$u(g_1, g_2) = \frac{\sum_k \min(F(g_1, o_k), F(g_2, o_k))}{\sum_k \max(F(g_1, o_k), F(g_2, o_k))} \qquad (6)$$

$$v(g_1, g_2) = \frac{\sum_k \min(F(g_1, o_k), F(g_2, o_k))}{\sum_k F(g_1, o_k)} \qquad (7)$$

The values *u(g$_1$, g$_2$)* and *v(g$_1$, g$_2$)* are obtained on the basis of the co-occurrences of goals $g_1$ and $g_2$ in the set O of Objects.

### 4.4. *Fuzzy clustering of objects*

Clustering in Knowledge Extraction is a method for partitioning a given set of Objects O into groups using a measure of similarity that is defined on every pairs of Objects. Similarity between Objects in the same group should be large, while it should be small for Objects in different groups. Generated clusters can then be used as an index for knowledge extraction; that is, also the Objects that belong to the same clusters of the Objects directly indexed by the goals in the request are extracted. Often, similarity measures are suggested empirically or heuristically [7].

When adopting fuzzy set theory, clustering can be formalised *as* a kind of fuzzy association. In this case, the fuzzy association is defined as f: O *x* O → [0, 1], where O is the set of Objects.

By assuming *R(o)* to be the fuzzy set of goals representing an Object O with membership function $\mu_o$ whose values $\mu_o(g) = F(o, g)$ are the index goal weights of goal g in Object o, the symmetric fuzzy relation *u, as* originally defined above, is taken to be the similarity measure for clustering Objects.

In fuzzy clustering, Objects can belong to more than one cluster with varying degree of membership. Each Object is assigned a membership value to each cluster. In a pure fuzzy clustering, a complete overlap of clusters is allowed.

Several researchers have worked on fuzzy clustering for Extraction, who include as Kamel, et al. [37], and Miyamoto [20], and De Mantaras et al. [38]. Modified fuzzy clustering, or clustering, approaches use thresholding mechanisms to limit the number of

Objects belonging to each cluster. The main advantage of using modified fuzzy clusturing is the fact that the degree of fuzziness is controlled.

## 5. Conclusions and Future Works

In this paper, a synthetic survey of some approaches to define flexible KESs has been presented. This survey refers to KE models based on the so-called Computing paradigm; in particular the approaches based on Fuzzy Set Theory and on Semantic Networks were considered to describe the Goals and Objects. The analysed approaches make it possible to model aspects of the inherent vagueness and uncertainty characterising the Knowledge Extraction process. The application of computing to KE is particularly appealing to model mechanisms which learn the user's notion of Objects' relevance. Future directions of this research area include the integration of neuro-fuzzy approaches and the definition of more powerful associative mechanism to improve the effectiveness of KESs.